\documentclass[fleqn,usenatbib]{mnras}

\usepackage{newtxtext,newtxmath}
\usepackage[T1]{fontenc}

\DeclareRobustCommand{\VAN}[3]{#2}
\let\VANthebibliography\thebibliography
\def\thebibliography{\DeclareRobustCommand{\VAN}[3]{##3}\VANthebibliography}

\usepackage{graphicx}	
\usepackage{amsmath}	
\usepackage{indentfirst}
\usepackage{url}  
\usepackage{booktabs}
\usepackage{caption}
\usepackage{subcaption}
\usepackage{colortbl}
\usepackage{hhline}
\usepackage{hyperref}
\usepackage{ragged2e}

\usepackage{soul} 
\usepackage{textcomp}

\numberwithin{equation}{section}

\hypersetup{
    unicode=false, pdftoolbar=true, 
    pdfmenubar=true, pdffitwindow=false, pdfstartview={FitH}, 
    pdftitle={ELE2024 Coursework}, pdfauthor={A. Author},
    pdfsubject={ELE2024 coursework}, pdfcreator={A. Author},
    pdfproducer={ELE2024}, pdfnewwindow=true,
    colorlinks=true, linkcolor=blue,
    citecolor=cyan, filecolor=magenta, urlcolor=cyan
}

\title[The Brazilian PHEMU21 Campaign]{The 2021 mutual phenomena involving the Galilean satellites of Jupiter and the inner satellite Thebe\thanks{Based in part on observations made at the Laborat\'orio Nacional de Astrof\'isica (LNA), Itajub\'a-MG, Brazil.}}

\author[L. M. Catani et al.]{
L. M. Catani,$^{1,2}$\thanks{E-mail: lcatani@ov.ufrj.br}
M. Assafin,$^{1,2}$
B. E. Morgado,$^{1,2}$
S. Santos-Filho,$^{1,2, 6}$
F. Braga-Ribas,$^{4}$
R. Vieira-Martins,$^{2,3}$
\newauthor
J. Arcas-Silva,$^{2,3}$
A. C. Milone,$^{5}$
I. J. Lima$^{5}$ and
R. B. Botelho$^{5}$
\\
$^{1}$Universidade Federal do Rio de Janeiro (UFRJ), Ladeira do Pedro Antonio 43, Rio de Janeiro, RJ 20080-090, Brazil\\
$^{2}$Laboratório Interinstitucional de e-Astronomia (LIneA), Rua Gal. José Cristino 77, Rio de Janeiro, RJ 20921-400, Brazil\\
$^{3}$Observatório Nacional/MCTI (ON), Rua General José Cristino 77, Rio de Janeiro, RJ 20921-400, Brazil\\
$^{4}$Federal University of Technology - Paraná (PPGFA/UTFPR-Curitiba), Av. Sete de setembro 3165, Curitiba, PR 80230-901, Brazil \\
$^{5}$Instituto Nacional de Pesquisas Espacias (INPE), Av. dos Astronautas, 1.758, São José dos Campos, SP 12227-010, Brazil \\
$^{6}$Centro de Educação a Distância do Estado do Rio de Janeiro (CEDERJ), Praça Cristiano Ottoni S/N – 6º andar, Rio de Janeiro, RJ 20221-250, Brazil
}

\date{Accepted XXX. Received YYY; in original form ZZZ}

\pubyear{2023}

\begin{document}
\label{firstpage}
\pagerange{\pageref{firstpage}--\pageref{lastpage}}
\maketitle

\begin{abstract} 

\noindent Astrometric studies and orbital modeling of planetary moons have contributed significantly to advancing our understanding of their orbital dynamics. These studies require precise positions measured over extended periods. In this paper, we present the results of the 2021 Brazilian Jovian mutual phenomena campaign. The data correspond to eight events between Galilean satellites, in addition to a rare eclipse of Thebe, an inner satellite, totaling nine events. A geometric model along with the DE440/JUP365 ephemerides was used to reproduce the events and simulate the light curves. A Monte Carlo method and chi-squared statistics were used to fit the simulated light curves to the observations. The reflectance model adopted for our simulations was the complete version of the Oren-Nayer model. The average uncertainty of the relative positions of the Galilean satellites was 5 $mas$ (15 km) and for the inner Thebe satellite 32 $mas$ (96 km). The seven mutual events (nine independent observations) here analyzed represent and addition of 17\% events (10\% light curves) with respect to the PHEMU21 international campaign. Furthermore, our result of Thebe eclipse is only the second measurement published to date. Our results contribute to the ephemeris database, being fundamental to improving satellite orbits and thus minimizing their uncertainties.  

\end{abstract}




\begin{table*}
	\centering
	\caption{Observational campaign details.}
	\label{tab:01}
	\begin{tabular}{lccccccc} 
		\hline
            \hline
		City/Country & Station abbreviation & Telescopes & Longitude & Altitude &  Observer & CCD & Nº of positive  \\
            \cmidrule(lr){2-2}
            \cmidrule(lr){3-3}
            \cmidrule(lr){4-4}
            & MPC code & Aperture & Latitude  &  &  & & detections\\
		
            \hline
		Itajubá/MG, & OPD & Boller \& Chivens & 45º34'57''W & 1864m & L. M. Catani & Andor-iXon$^{EM}$ & 6\\
            Brazil & 874 & 0.60m & 22º32'07''S &  & B. E. Morgado & &\\
            \cmidrule{3-3}
            & & Perkin-Elmer & & & S. Santo-Filho & &\\
            & & 1.60m & & & J. Arcas-Silva & &\\
            & & & & & A. R. Gomes-Junior & &\\
            & & & & &  \\
		Curitiba/PR, & UTFPR & Meade-LXD55 & 49º20'23''W &  935m & F. Braga-Ribas & QHY174MGPS & 2 \\
            Brazil & -- & 0.20m & 25º26'10''S & & & &\\
            & & & & &  \\
		São José-  & INPE & Celestron-C11 & 45º51'43''W & 620m  & A. C. Milone & Watec 910HX & 1\\
            dos Campos/SP, & -- & 0.28m  & 23º12'32''S & & R. B. Botelho & &\\
            Brazil & & & & & I. J. Lima & &\\
		\hline
            
	\end{tabular}
\end{table*}

\section{Introduction}

The study of the orbital dynamics of satellite systems around giant planets is crucial for enhancing our understanding of the formation and evolution of the Solar System \citep{arlot2007}. While the orbital evolution of these systems resembles that of objects in solar orbit, it occurs on a smaller time scale at a considerably faster rate. Consequently, investigating the physical aspects and dynamic evolution of planetary satellite systems becomes essential as it enables the examination of dynamic perturbations, including resonant and secular effects resulting from tidal dissipation \citep{lainey2009}. Apart from shedding light on the system's dynamic evolution, these measurements are key in determining more accurate positions and updating their ephemerides.

In the Jovian system, there is a group of satellites located within the orbits of the Galilean moons. These small moons have a particular interest as they are believed to be the source of particles to Jupiter's ring system \citep{ockert1999}. Studying the kinematics of these satellites helps us better understand the structure of Jupiter's rings and how they are maintained. However, due to the difficulty of observing these objects from ground-based observatories, the astrometry of these objects is particularly challenging \citep{veiga1994}. 

The astrometric positions of Jupiter's main and inner satellites can be determined by observing mutual phenomena. These phenomena occur when one natural satellite eclipses or occults another from the perspective of an observer on Earth. Such events occur when the Earth and the Sun align with the common orbital plane of Jupiter's satellites. This particular alignment happens twice during each planet's orbit around the Sun. In the case of Jupiter, these alignments occur every six years, the last one being in 2021.

During mutual phenomena, we use photometry to measure object brightness variations over time. This technique precisely determines the relative positions between the satellites involved in the event \citep{arlot2019, emelyanov2009, emelyanov2022}. The astrometry with this method typically has an uncertainty of only about 5 \textit{mas}, which is equivalent to about 15 kilometers at Jupiter's distance \citep{arlot2014, emelyanov2009, dias2013, morgado2019}. For example, classical CCD astrometry exhibits uncertainties around 100 \textit{mas} ($\sim$ 300 km) \citep{kiseleva2008}, mutual approximations present uncertainties of the order of 10 \textit{mas}  \citep[$\sim$ 30 km;][]{morgado2016, morgado2019approx}, astrometric measurements obtained via radio show minimum uncertainties around 0.5 \textit{mas} \cite[$\sim$ 1.5 km; ][]{brozovic2020} and the technique of stellar occultations achieved an uncertainty of 0.8 \textit{mas} \citep[$\sim$ 2.4 km; ][]{morgado2019occ,morgado2022} . In addition, for the inner satellites, the average accuracy achieved through mutual phenomena for Thebe was around 45 $mas$ \citep[$\sim$ 135 km; ][]{saquet2016}, and for Amalthea, it is 82 $mas$ ($\sim$ 246 km) as reported by \cite{christou2010} and 47.8 $mas$ ($\sim$ 143 km) given by \cite{morgado2019}.

This paper presents the results obtained from the analysis of the 2021 mutual phenomena of the Galilean satellites, including a rare eclipse involving the inner satellite Thebe (J14) of Jupiter. Mutual phenomena involving an inner satellite were first published by \cite{christou2010} in 2019. Still,
only one other event involving Thebe has been published so far by \cite{saquet2016}. Altogether, we analyzed nine mutual phenomena involving Jupiter's satellites.

In Section \ref{sec:02}, we provide the details of the observational campaign and the employed instrumentation. Section \ref{sec:03} presents the techniques used to mitigate the scattered brightness from Jupiter in the observed images with the I filter. The photometric reduction method of the observational data is addressed in Section \ref{sec:04}. Sections \ref{sec:5}, \ref{sec:6}, and \ref{sec:7} delve into the modeling specifics, the process of fitting light curves, and parameter conversion. The results obtained and the conclusion are presented in Sections \ref{sec:8} and \ref{sec:9}.


\begin{table}
    \centering
    \caption{Observed events. 
    \label{tab:02}}
    \begin{tabular}{cccccc}
    \hline \hline
{Date} & {Observer} & {Event} & {Central} & {$\Delta t$} & {Flux drop} \\ 
{aa/mm/dd} & {} &{($S_1 x S_2$)} & {(UTC)} & {(min)} & {} 
\\
    \hline 
21/04/10 & OPD & 3ecl4 & 07:20 & 17.9 & 0.089 \\
21/04/11 & OPD  & 1ecl2 & 07:39 & 5.0 & 0.626 \\
21/05/20 & OPD & 1ecl2 & 07:56 & 5.4 & 0.489 \\ 
21/06/21 & OPD & 1ecl2 & 06:12 & 4.1 & 0.074 \\ 
21/07/18 & OPD & \phantom{0}3ecl14 & 04:04 & 2.8 & - \\ 
21/08/09 & OPD  & 3occ2 & 06:17 & 29.1 & 0.157 \\ 
21/08/09 & UTFPR & 3ecl2 & 03:37 & 67.5 & 0.465 \\ 
21/08/09 & UTFPR & 3occ2 & 06:17 & 29.1 & 0.157 \\ 
21/08/09 & INPE& 3ecl2 & 03:37 & 67.5 & 0.465 \\ 

    \hline 
    \end{tabular}

    \noindent \justifying{\textit{Note:} Information about codes in the MPC and geographic locations of observation sites can be seen in the Table \ref{tab:01}. 
    }
    
\end{table}


\section{Observational Campaign}\label{sec:02}

In 2021, Jupiter entered its equinox. This provided an opportunity to observe mutual phenomena among the main Jovian satellites. The observations of the Brazilian mutual phenomena campaign in 2021 were carried out at the Pico dos Dias Observatory (OPD) and by observers in the South and South-East of Brazil. The predicted events were based on the ephemerides provided by the \textit{Institut de Mécanique Céleste et de Calcul des Ephémérides} (IMCCE).{\footnote{Website: \href{http://nsdb.imcce.fr/multisat/nsszph517he.htm}{http://nsdb.imcce.fr/multisat/nsszph517he.htm}}$^,$\footnote{Website: \href{http://nsdb.imcce.fr/multisat/nsszph518he.htm}{http://nsdb.imcce.fr/multisat/nsszph518he.htm}}} 

We have gathered in Table \ref{tab:01} the information about the observational campaign: observation locations, data collection station abbreviations, MPC codes (when available), telescope apertures, geographic coordinates, altitude, observers, the sensor used, and the number of positive detections.

\subsection{Selection of observed events}

During the 2021 campaign, Jupiter had declination from -16º to -12º, which favored observations in the Southern Hemisphere. 

A total of 192 mutual events were predicted for the Galilean satellites, visible for both hemispheres \citep{arlot2019}. Of the total predicted events, only 47 would be observable from Brazil, and out of these, only 37 were visible from the observation sites. We selected 12 events among the Galilean satellites to be observed in our campaign. Additionally, we chose 6 events involving the inner satellites, specifically Thebe and Amalthea.

The observed events are listed in Table \ref{tab:02}, which includes the dates of each event; the station; the type of event (ecl for eclipse and occ for occultations); and the satellites involved, 1 for Io, 2 for Europa, 3 for Ganymede, 4 for Callisto and 14 for Thebe. In addition, the table also displays the expected central instant of each event, the duration, and the predicted magnitude drop. Note that, for events involving the inner satellites, the IMCCE does not provide predictions of flux drop for the events.

Of the eighteen selected events, we obtained nine light curves for analysis. The nine light curves were generated from seven different events, two of these seven events were observed by more than one observer, totaling nine light curves.
The observed events are indicated in the third column of the Table \ref{tab:02}. Of these events, we observe an eclipse between the inner satellite Thebe and Ganymede.
The other events on the list were lost due to unfavorable weather conditions.

\subsection{Instrumentation}

The telescopes employed for observing the Galilean satellites had apertures ranging from 203 mm to 600 mm. For the observations of events involving the inner satellites, we specifically utilized the Perkin-Elmer 1.6 m telescope located at the Pico do Dias Observatory.

In observations involving only the Galilean satellites, we employed a narrowband filter with a central wavelength of 889 nm and a bandwidth of 15 nm. This specific band falls within the methane absorption region of the electromagnetic spectrum. We deliberately chose this filter because, in this spectral range, Jupiter's albedo decreases to 0.05 as a result of absorption caused by the presence of methane in its upper atmosphere \citep{karkoschka1998}. 

\begin{figure}
    \centering
    \vspace{0.1cm}
    \begin{minipage}{0.235\textwidth}
    \centering
    \includegraphics[width=1\textwidth]{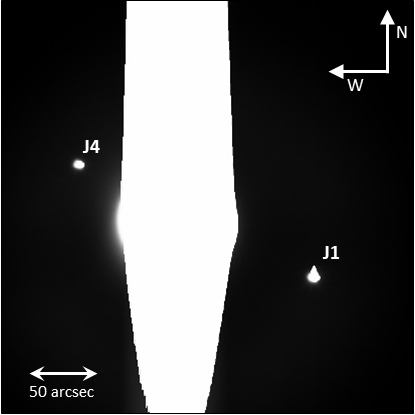}
    \subcaption{I Filter of Johnson System}
    \end{minipage}
    \hfill
    \centering
    \begin{minipage}{0.235\textwidth}
    \centering
    \includegraphics[width=1\textwidth]{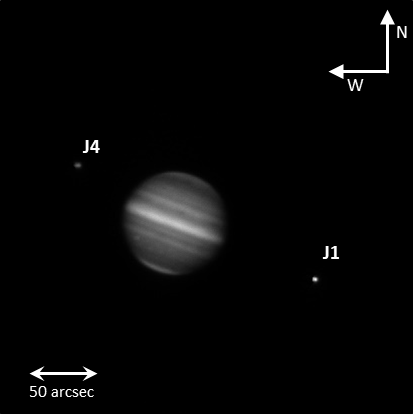}
    \subcaption{Narrow-band filter in 889 nm}
    \end{minipage}
    \caption{Jupiter observed the two filters used during the campaign. The images were taken on July 5, 2021, with the 1.6 m Perkin-Elmer telescope, both with a one-second exposure time.}
    \label{fig:01}
\end{figure}

We show in Fig. \ref{fig:01}\textcolor{blue}{b} an image obtained using the Perkin-Elmer 1.6m telescope at the OPD, providing an illustrative example of the methane filter effect. The planet and the satellites present about the same brightness due to the use of the narrow-band filter. This filter has been successfully used in the mutual phenomena campaigns of 2009 and 2014-2015 \citep{dias2013, morgado2019}, as well as in the mutual approximation campaigns initiated in 2014 \citep{morgado2016, morgado2019approx}.

On the other hand, for observations of events involving the inner satellites, we used the I filter from the Johnson system, shown in Fig. \ref{fig:01}\textcolor{blue}{a}. We opted for this filter because it allows more light to pass through compared to the methane filter, favoring the capture of the faint brightness of the inner satellites and avoiding the complete saturation of the CCD, which would happen if no filter were used in the observation of Jupiter with a large telescope. In this context, it was necessary to apply additional processing to minimize the contribution of Jupiter's luminosity in the images, as explained in the next section.

\section{Treatment of the images obtained with an I filter}\label{sec:03}

In the images we observe using the I filter, we apply a technique to reduce part of the contamination of the scattered-light halo surrounding Jupiter over the inner satellites. This technique masking scattered light from Jupiter in the image. We construct a Jupiter model to subtract from the observed data. We construct the template using different sets of images from the same observation. In this context, when applying them to science images, it is necessary to take certain precautions to preserve the integrity of the scientific data. Some of these precautions include ensuring that the flux of the target object is not subtracted and ensuring that the telescope guiding is accurate enough to avoid significant variations in the observed field, as the effectiveness of this method relies on minimizing the movement of the contaminating source in the images.

\begin{figure*}
    \centering
    \begin{minipage}{0.327\textwidth}
    \centering
    \includegraphics[width=1\textwidth]{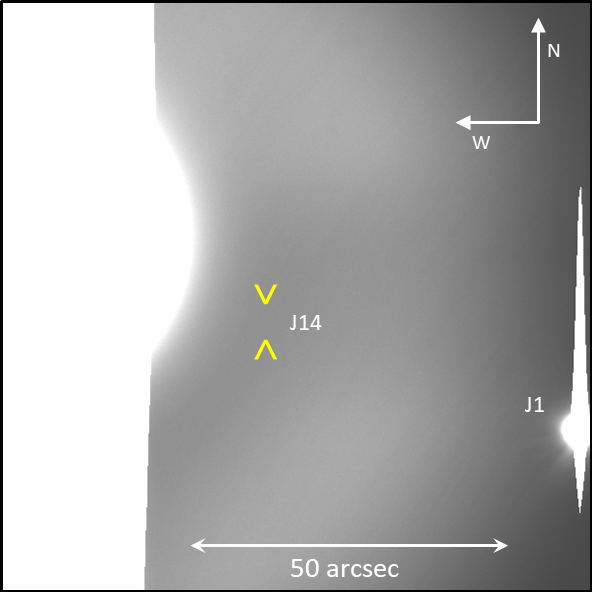}
    \subcaption{Standard calibration}
    \end{minipage}
    \hfill
    \centering
    \begin{minipage}{0.327\textwidth}
    \centering
    \includegraphics[width=1\textwidth]{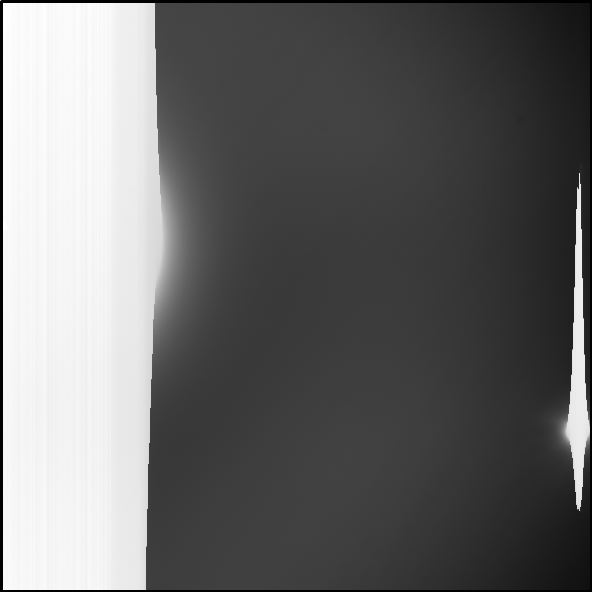}
    \subcaption{Template}
    \end{minipage}
    \hfill
    \centering
    \begin{minipage}{0.327\textwidth}
    \centering
    \includegraphics[width=1\textwidth]{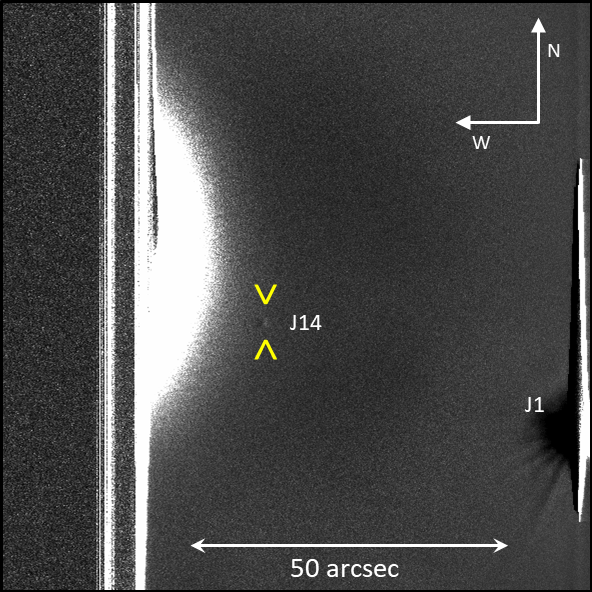}
    \subcaption{After subtraction with template}
    \end{minipage}
    \caption{\justifying{Result obtained with the template technique. In (a) the image with standard treatment of \textit{bias} and \textit{flat-field}; in (b) the \textit{template master} constructed for that set; and in (c) the image after applying the technique. The yellow arrow indicates Thebe's position in the image. This event was the eclipse between Ganymede (J3) and Thebe (J14) observed on July 18, 2021, by the Perkin-Elmer treaty applying 5 seconds of exposure using the I filter of the Johnson system.}}
    \label{fig:03}
\end{figure*}

The application of this technique is particularly useful for the inner satellites of Jupiter, which have a slightly faster orbital period, resulting in a significant displacement in the image field over short intervals of time. The result of this application can be seen sequentially in Fig. \ref{fig:03}, where we have an original image with standard calibration (Fig. \ref{fig:03}\textcolor{blue}{.a}), the template with the measured luminous contribution from the observations (Fig. \ref{fig:03}\textcolor{blue}{.b}), and in Fig. \ref{fig:03}\textcolor{blue}{.c}, the image after the applying this approach.

Note that this technique was applied after the standard bias and flat-field calibration, a procedure performed with the IRAF software \citep{butcher1981}, the same used for the application of the technique detailed here.

\begin{figure}
    \centering
	\includegraphics[width=\columnwidth]{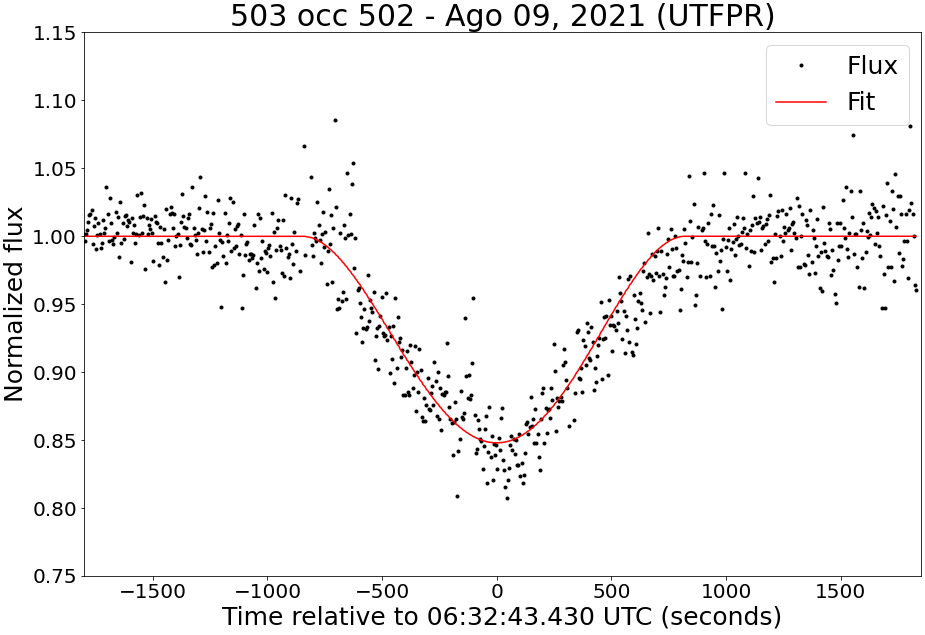}
    \vspace{-0.5cm}
    \caption{ Light curve of an occultation between Ganymede (503) and Callisto (504) observed by UTFPR collaborators on August 09, 2021. The red line represents the modeled flux of the event (see Section \ref{sec:5}).}
    \label{fig:02}
\end{figure}

\section{Photometric Reduction}\label{sec:04}

The astrometric study of mutual phenomena is conducted by analyzing the light curves obtained from the observation of these events. Here, the light curves were generated with differential aperture photometry. This approach is particularly useful when you want to measure subtle variations in an object's luminous flux (magnitude) over time \citep{kjeldsen1992}. To build the light curves from the observations, we used the tasks of PRAIA \citep[Package for the Reduction of Astronomical Images Automatically; ][]{assafin2023}, which facilitated the analysis and processing of the data. The light curve is then normalized by a polynomial fit so that the flux ratio outside the flux drop gets equal to 1.0, and the flux drop can be adequately evaluated.

We present an example of light curves obtained with PRAIA in Fig. \ref{fig:02}. The dots illustrate the normalized measurements of light flux, while the red line represents the modeled flux of the event, predicted by the DE440 + JUP365 ephemeris and adjusted based on the observations (see details in Section \ref{sec:5}). 

The light flux determination in each image was achieved through aperture photometry, integrating all the light flux within a specific area. This approach effectively eliminates atmospheric and sky background effects by referencing a calibration object. During an occultation, both satellites are measured together in the same aperture, and a third satellite is used as a calibrator. In the case of eclipses, the eclipsed satellite is measured alone in the aperture, and the eclipsing satellite (or any other) is used as a calibrator.

In Table \ref{tab:03}, we have gathered information regarding the photometric reduction of the nine light curves derived from our observations. In addition to the analogous information from Table \ref{tab:02}, we display the calibration object used in the fourth column, the aperture size in pixels employed during the reduction in the fifth column, and the sigma values of the flux ratio in the last column. Note that for two events, we lack a calibration object. Therefore, the high frequency variations in flux for these events are a result of atmospheric effects that could not be eliminated.

\section{Method for analyzing the light curve and obtaining astrometric data}\label{sec:5}

Here we will discuss the method used to analyze light curves obtained from the observation and obtain astrometric data. Through the analysis of light curves from mutual phenomena, we can make precise measurements of the relative positions of the satellites involved in the event.

During the analysis of the curves, we seek to relate three parameters for the calculation of satellite positions. The parameters of interest in the analysis include the impact parameter ($S_0$), which is related to the smallest apparent angular distance between the satellites in the sky plane. In an occultation, this parameter refers to the minimum apparent distance between the satellites. In the case of an eclipse, it is related to the minimum apparent separation between the geometric centers of the eclipsed satellite and the shadow of the eclipsing satellite in the sky plane. Another parameter is the central instant ($t_0$), which refers to the exact instant when the smallest apparent distance between the satellites occurs. Additionally, the analysis also takes into account the apparent relative velocity ($v_0$) between the two satellites in the celestial plane.

These parameters are intrinsically linked to the dynamics and geometry of the satellites involved in the event. As a result, the value of each parameter has a direct relationship with the appearance of the light curves of the events. In this context, the analysis method used in this study involves the simulation of theoretical light curves for different values of each parameter. These theoretical curves are then compared to the light curves obtained from the observation, seeking the one that best fits the experimental data. This approach allows for the precise determination of parameter values and the obtaining of relative satellite positions.

\begin{table}
    \centering
    \caption{Information about photometric reduction.  
    \label{tab:03}}
    \begin{tabular}{cccccc}
    \hline \hline
{Date} & {Observer} & {Event} & {Calib.} & {Aperture} & {$\sigma$ of flux} \\ 
{aa/mm/dd} & {} &{} & {} & {in pixels} & {ratio} 
\\
    \midrule
21/04/10 & OPD & 3ecl4 & 3 & 11 & 0.019 \\
21/04/11 & OPD & 1ecl2 & 1 & 13 & 0.018 \\
21/05/20 & OPD & 1ecl2 & 1 & 10 & 0.011 \\ 
21/06/21 & OPD & 1ecl2 & 1 & 9 & 0.006 \\ 
21/07/18 & OPD & \phantom{0}3ecl14 & - & 5 & 0.210 \\ 
21/08/09 & OPD & 3occ2 & - & 6 & 0.017 \\ 
21/08/09 & UTFPR & 3ecl2 & 3 & 2 & 0.043 \\ 
21/08/09 & UTFPR & 3occ2 & 1 & 25 & 0.001 \\ 
21/08/09 & INPE & 3ecl2 & 3 & 4 & 0.126 \\

    \hline
    \end{tabular}
    
    \noindent \justifying{\textit{Note:} Here, analogously to Table \ref{tab:02}, we have: 1 = Io, 2 = Europa, 3 = Ganymede, 4 = Callisto, and 14 = Thebe.}
    
\end{table}


\subsection{Geometric model for the theoretical light curves}

It is crucial to select an appropriate model to generate theoretical light curves, as the accuracy of simulated light curves depends on the chosen model for their construction. The geometric model we use to simulate theoretical light curves follows the principles described in \cite{assafin2009}, \cite{dias2013}, and \cite{morgado2019}.

The adopted model simulates the light curve based on six parameters, which can be divided into two categories: physical parameters and dynamic parameters. The physical parameters of the model are the apparent sizes and shapes of each satellite and the albedo of the satellites in the case of occultations. The dynamic parameters are the ones we aim to obtain through the analysis, namely, the parameters of interest: $S_0$, $t_0$, and $v_0$.

The physical parameters, such as radius and shape, are naturally kept fixed during the modeling since they are determined with high precision from data collected by space probes. The albedos, on the other hand, are measured through auxiliary observations conducted before and after the events using the same instrumental configuration. In some specific cases where it is not possible to obtain albedo measurements on the same night, previously determined albedo values are used.

On the other hand, unlike the physical parameters, the dynamic parameters are not fixed. By adopting different values for these parameters, it is possible to modify the appearance of simulated light curves. Based on this, a large number of theoretical light curves are generated using different sets of values for each parameter of interest. The curve that best matches the observed curve is selected. The fitting procedure follows the statistical method of chi-square minimization, as detailed in Section \ref{sec:6}.  

\subsubsection{Simulating eclipses and ocultations}

Now let's consider the differences between the simulation of eclipses and occultations. In general terms, the geometric model developed by \cite{dias2013} and refined by \cite{morgado2019} constructs two-dimensional profiles based on the input parameters, simulating how the bodies are seen during these events by an observer on Earth.

To simulate an eclipse, the model constructs the disk of the eclipsed satellite and the shadow cast by the eclipsing satellite. A numerical model is used to reproduce the penumbra effect. In the case of occultations, instead of the shadow profile, the model reproduces the disks of the occulting and occulted satellites.

We used a reflectance model to simulate the reflection of light from the surface of the satellites with greater accuracy, as well as the effects of the solar phase. Following the steps of \cite{morgado2019}, we employed the reflectance model proposed by \cite{oren1994}, which utilizes a general law for the reflection of non-homogeneous disks.

In both cases, eclipses and occultations, it is necessary to know the albedo of the satellites in the specific wavelength at which the observation is conducted. This is crucial to accurately represent the profile of each satellite exactly as they are observed.

In the case of occultations, it is necessary to calculate the albedo ratio since the measurement of the flux from both satellites is done using the same aperture. To do this, we measure the flux ($F_1, F_2$) of both satellites separately, using the same instrumental setup. It is preferable to perform these measurements on the same night the event was observed.

The albedo ratio is then calculated using equation (\ref{eq:2}), which provides a relationship between the observed fluxes ($F_1, F_2$) with albedos ($A_1, A_2$) and the modeled fluxes ($F_{S1}, F_{S2}$).

\begin{equation}\label{eq:2}
    \frac{F_1}{F_2} = \frac{A_1}{A_2} \cdot \frac{F_{s1}}{F_{s2}}.
\end{equation}

Another important aspect is related to the physical parameters of the Sun, such as its radius, as well as using a model that considers limb darkening. The limb darkening effect was modeled according to the law of \cite{hestroffer1998}, with the parameters defined according to the spectral region of the observation.

Note that both limb darkening and reflectance are crucial elements for obtaining an accurate representation of the simulated light curves.

The final step in generating the light curves involves associating the modeled profiles with the actual observations. The code accomplishes this by simulating the flux F(t) of the satellites, integrating the modeled profile over the exposure times for each observation until the simulated light curve is obtained. Finally, the light curve is normalized outside of the event time window with respect to the flux of the eclipsed or occulted satellite, providing the simulated light curve.

\section{Fitting Light Curves models to Observations curves}\label{sec:6}

The fitting of the light curves involves comparing the observed light flux ($f_i$) with the modeled light flux ($f'_i$) for each image ($i$) using the chi-squared method, as presented in equation (\ref{eq:3}).

\begin{equation}\label{eq:3}
    \chi^2 = \sum_{i = 1}^{N} \left ( \frac{(f_{i} - f'_{i})}{\sigma} \right )^2.
\end{equation}

\noindent Where $\sigma$ was obtained by calculating the standard deviation of the observed light curve in the linear region outside the event.

The marginal uncertainties for a confidence level of 68\% (1 sigma) are determined from the parameters that result in $\chi^2$ values less than $\chi^2_{\text{min}} + 1$. For a good fit, it is expected that $\chi^2$ is approximately equal to the degrees of freedom, which is defined as the difference between the number of data points ($N$) and the number of model parameters ($P = 3$).

Based on the adopted model, a large number of artificial light curves (about nine hundred for each event) are generated by varying the values of each dynamic parameter within a specific range. This range of values revolves around the predicted values for each event according to the DE440 + JUP365 ephemerides. Subsequently, corrections are applied to each light curve in order to find the parameters that minimize the chi-squared test. This procedure is performed using a Python code developed by \cite{morgado2019}.

\begin{table*}
    \centering
    
    \caption{Results of the parameters of interest obtained for the Galileans.}\label{tab:04}
    \begin{tabular}{cccccccccccccc}
    \hline \hline
{Date} & {Event} & Obs. & {$S_0$} & {Error ($S_0$)} & {$t_{0}$ UTC} & {Error ($t_0$)} & {$v_0$} & {Error ($v_0$)} & {$\Delta S_0$} & {$\Delta t_0$} & {$\Delta v_0$}  & $N$ & $\chi^2$ 
\\ 
{(y/m/d)} & {$S_1 x S_2$} & {} & {($mas$)} & {($mas$)} & {(h:m:s.s)} & {($s$)} & {($mas/s$)} & {($mas/s$)} & {($mas$)} & {($mas$)} & {($mas$)}  & {} & {}
\\
\hline 

21/04/10 & 3ecl4 & OPD & 1331.8 & 3.57 & 07:30:29.33 & 1.06 & 2.31 & 0.10  & -3.48 & -2.72 & +0.03 & 242 & 0.538 \\
21/04/11 & 1ecl2 & OPD & 295.9 & 1.25 & 07:41:37.81 & 0.44 & 7.36 & 0.22 & -0.96 & -1.36 & -0.10 & 297 & 0.811 \\
21/05/20 & 1ecl2 & OPD &  466.9 & 1.54 & 07:58:40.32 & 0.45 & 6.31 & 0.50 & +1.89 & -0.15 & -0.10 & 169 & 0.700\\ 
21/06/21 & 1ecl2 & OPD & 908.5 & 10.8 & 06:14:38.47 & 0.92 & 5.28 & 0.79  & -10.81 & +2.10 & +0.06 & 86 & 0.609\\
21/08/09 & 3ecl2 & UTFPR & 414.3 & 3.68 & 04:13:38.90 & 6.45 & 0.61 & 0.05 & -11.60 & +2.09 & +0.01 & 717 & 0.323\\
21/08/09 & 3ecl2 & INPE &  413.2 & 9.17 & 04:13:43.27 & 18.11 & 0.62 & 0.05 & -12.80 & +4.98 & -0.02 & 1645 & 0.521 \\
21/08/09 & 3occ2 & OPD & 958.3 & 3.17 & 06:32:43.79 &3.18 & 1.21 & 0.07  & -4.10 & +0.44 & -0.02 & 172 & 0.859 \\
21/08/09 & 3occ2 & UTFPR & 958.3 & 4.46 & 06:32:44.25 & 3.47 & 1.23 & 0.08 & -3.04 & +1.29 & -0.01 & 250 & 0.741 \\

\hline 

{Mean} & {} & {} & {} & {4.70} & {} & {4.61} & {} & {0.23} & -5.61 & +0.83 & +0.01 & {} & {}\\
{S. D.}  & {} & {} & {} & {3.24} & {} & {2.66} & {} & {0.25} & 5.07 & 2.21 & 0.05 & {} & {} \\

\hline 
\end{tabular}


\end{table*}

\section{Converting parameters to $X$ and $Y$}\label{sec:7}

In our analysis, we obtain values for the impact parameter ($S_0$), central instant ($t_0$), and relative velocity ($v_0$), along with their corresponding offsets ($\Delta S_0$, $\Delta t_0$, and $\Delta v_0$) with respect to the ephemerides. Using these results, we can calculate the apparent separation of the satellites in the sky plane following equation (\ref{eq:4}) introduced by \cite{assafin2009}. This approach serves as the theoretical basis for mutual phenomena and mutual approximations techniques \citep{morgado2016, santos2019, morgado2019approx}.

\begin{equation}\label{eq:4}
    S(t) = \sqrt{S_{0}^{2} + v_{0}^{2}(t -t_0)^2}.
\end{equation} 

On the other hand, we can express them in terms of right ascension ($X = \Delta \alpha \cdot \cos(\delta)$) and declination ($Y = \Delta \delta$) following \cite{santos2019}. Here, $\delta = (\delta_1 + \delta_2)/2$, $\Delta \alpha = (\alpha_1 - \alpha_2)$, and $\Delta \delta = (\delta_1 - \delta_2)$, where index 1 refers to the occulting/eclipsing satellite 1, and index 2 to the occulted/eclipsed satellite 2. Ephemeris offsets can be obtained using equation (\ref{eq:5}). Where $\theta$ represents the position angle of satellite 1 relative to satellite 2 in the plane of the sky (in the counterclockwise direction, with zero in the eastward direction), this angle is calculated from the topocentric positions of the satellites. Additionally, by convention, the relative velocity ($v_0$) is defined as negative/positive when it points towards an increase/decrease in $\theta$ at $t_0$.

\begin{equation}\label{eq:5}
\begin{aligned}
    \Delta X = \Delta S_0 \cos(\theta) - \Delta t_0 \cdot v_0 \sin(\theta)  \\
    \Delta Y = \Delta S_0 \sin(\theta) + \Delta t_0 \cdot v_0 \cos(\theta). 
\end{aligned}
\end{equation}

The observed relative positions in $X$ and $Y$ can be calculated from the known relative distances obtained from the ephemerides and the values of ($\Delta X$, $\Delta Y$). The uncertainties in ($X$, $Y$) are determined from the errors in $t_0$ and $S_0$.

\section{Results of 2021 Jupiter's Mutual Phenomena}\label{sec:8}

The results of the mutual phenomena of Jupiter from the 2021 campaign were divided into two subsections. In the first one, we presented the analyses of the events that involved only the Galilean satellites. In the second one, we showed the results of the eclipse of the inner satellite, Thebe.


\begin{table*}

\begin{center}
\caption{Results in $X$ and $Y$ for events among Galileans.}\label{tab:05} \begin{tabular}{cccccccccc}
\hline \hline
    
{Date} & {Event} & {Obs.} & {$X$} & {$Y$} & {$\sigma X$} & {$\sigma Y$} & {$\Delta X$} & {$\Delta Y$} & {$N$} \\ 
{(y/m/d)} & {$S_1 x S_2$} & {} & {($mas$)} & {($mas$)} & {($mas$)} & {($mas$)} & {($mas$)} & {($mas$)} & {}
\\

\hline 

21/04/10 & 3 ecl 4 & OPD & +444.3 & -1255.5 & 3.47 & 2.57 & +1.42 & -4.19 & 242 \\
21/04/11 & 1 ecl 2 & OPD & -97.6 & +280.4 & 3.47 & 0.21 & +0.96 & -1.36 & 297 \\
21/05/20 & 1 ecl 2 & OPD & +163.6 & -437.3 & 2.17 & 2.46 & +0.80 & +1.72 & 169 \\ 
21/06/21 & 1 ecl 2 & OPD & +326.8 & -848.2 & 0.58 & 11.83 & -5.88 & -9.32 & 86\\
21/08/09 & 3 ecl 2 & UTFPR & -182.6 & +397.7 & 2.07 & 4.99 & -6.70 & -9.70 & 717\\
21/08/09 & 3 ecl 2 & INPE & -185.6 & +397.7 & 6.30 & 12.92 & -9.82 & -9.60 & 1645\\
21/08/09 & 3 occ 2 & OPD & -402.8 & +876.6 & 2.26 & 4.51 & -2.12 & -3.54 & 172\\
21/08/09 & 3 occ 2 & UTFPR & -402.7 & +875.2 & 2.02 & 5.87 & -2.46 & -2.22 & 250 \\

\hline 

{Mean}  & {} & {} & {} & {} &  2.79 & 5.67 & -2.97 & -4.77 \\
{S. D} & {} & {} & {} & {} &  1.57 & 4.21 & 3.86 & 4.03 \\

\hline 

\end{tabular}
\end{center}
\end{table*}


\begin{figure*}
    \centering
    \vspace{0.1cm}
    \begin{minipage}{0.44\textwidth}
    \centering
    \includegraphics[width=1\textwidth]{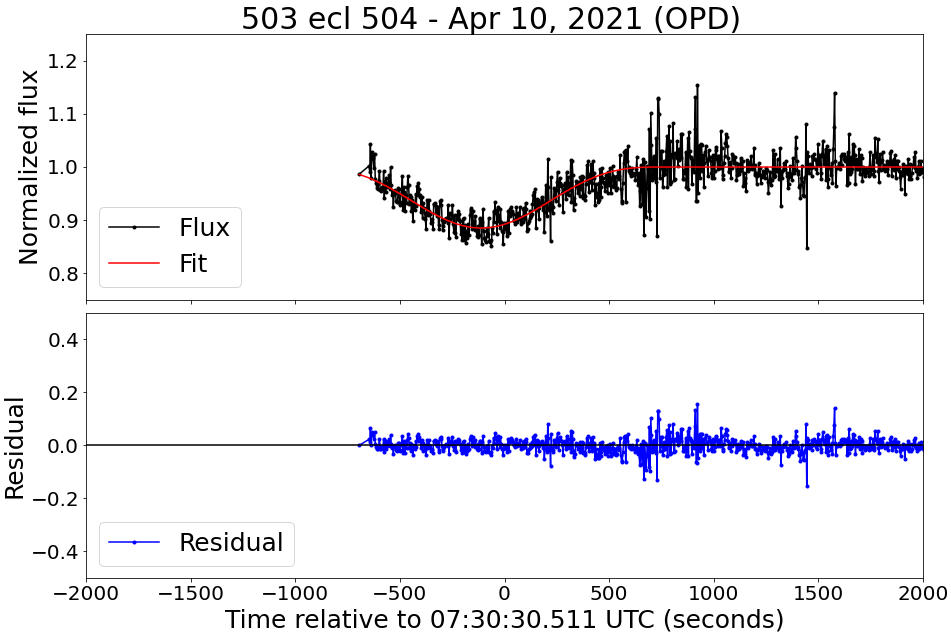}

    \end{minipage}
    \hspace{0.85cm}
    \centering
    \begin{minipage}{0.44\textwidth}
    \centering
    \includegraphics[width=1\textwidth]{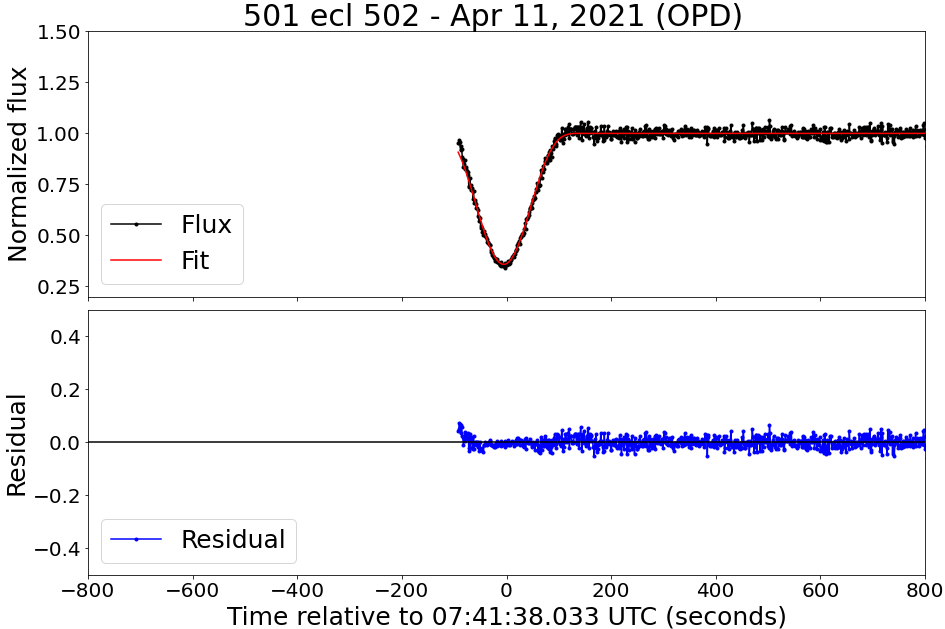}

    \end{minipage}
    \vfill
    \vspace{0.40cm}
    \centering
    \begin{minipage}{0.44\textwidth}
    \centering
    \includegraphics[width=1\textwidth]{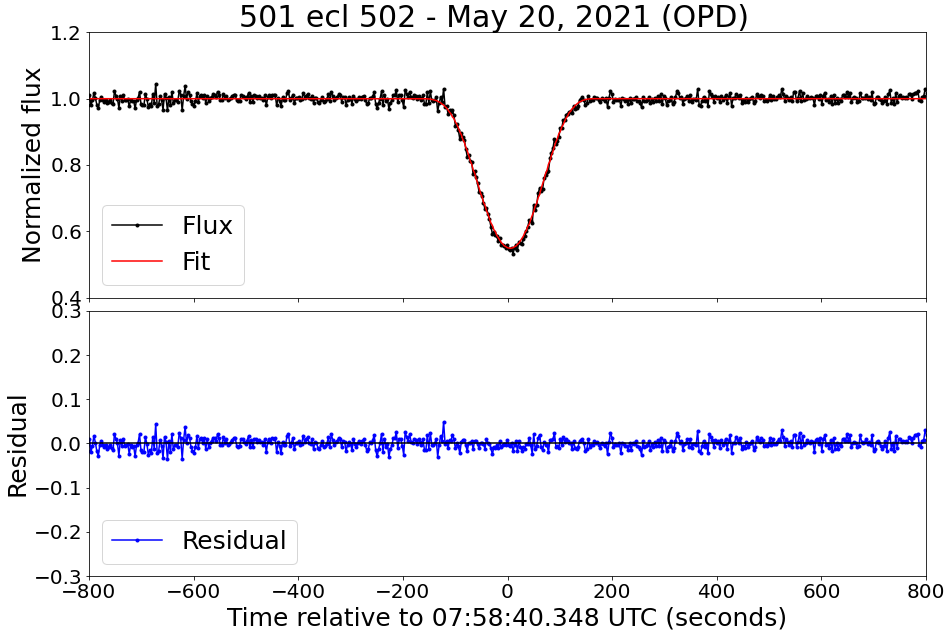}

    \end{minipage}
    \hspace{0.85cm}
    \centering
    \begin{minipage}{0.44\textwidth}
    \centering
    \includegraphics[width=1\textwidth]{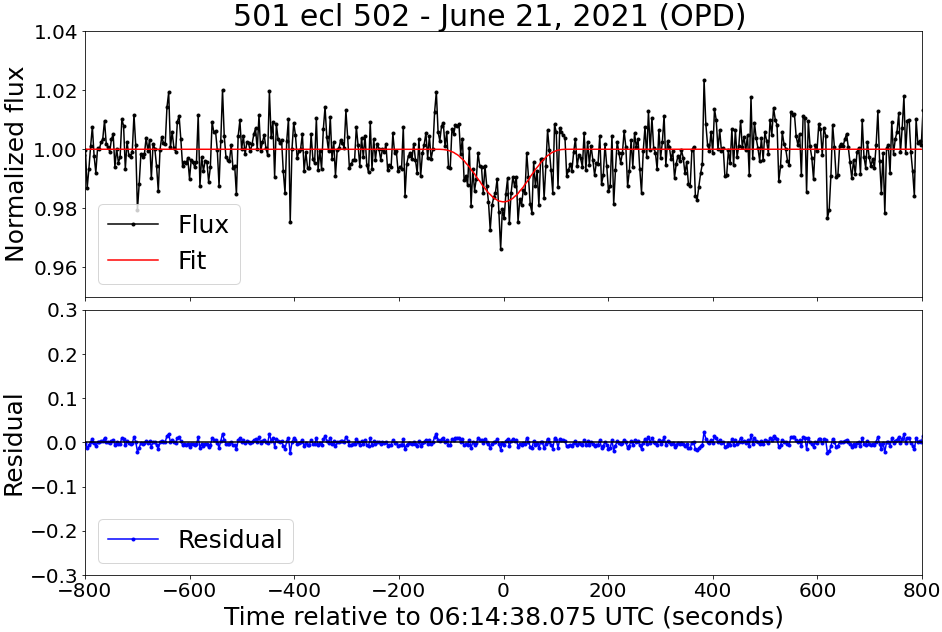}

    \end{minipage}
    \vfill
    \vspace{0.40cm}
    \centering
    \begin{minipage}{0.44\textwidth}
    \centering
    \includegraphics[width=1\textwidth]{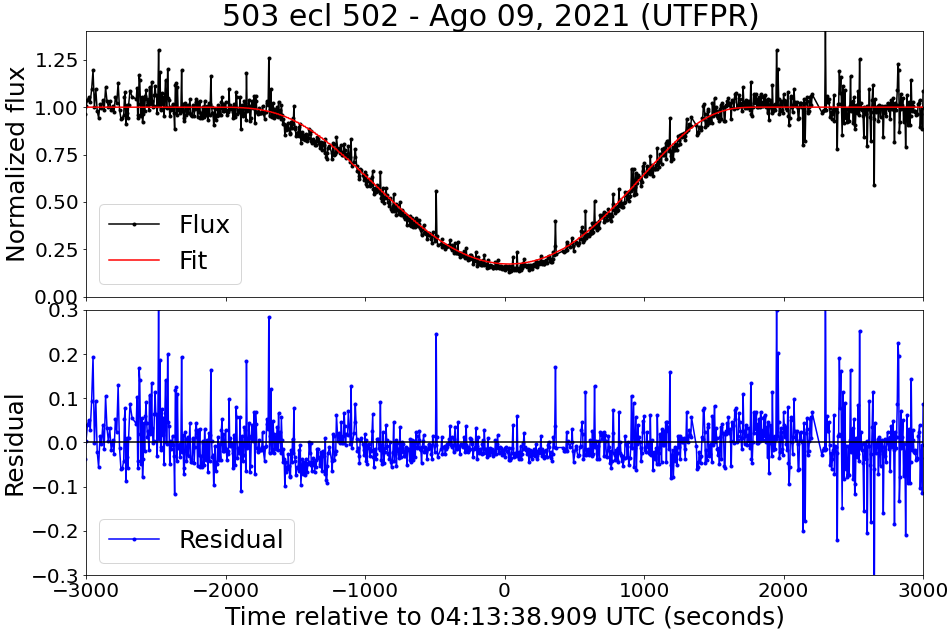}

    \end{minipage}
    \hspace{0.85cm}
    \centering
    \begin{minipage}{0.44\textwidth}
    \centering
    \includegraphics[width=1\textwidth]{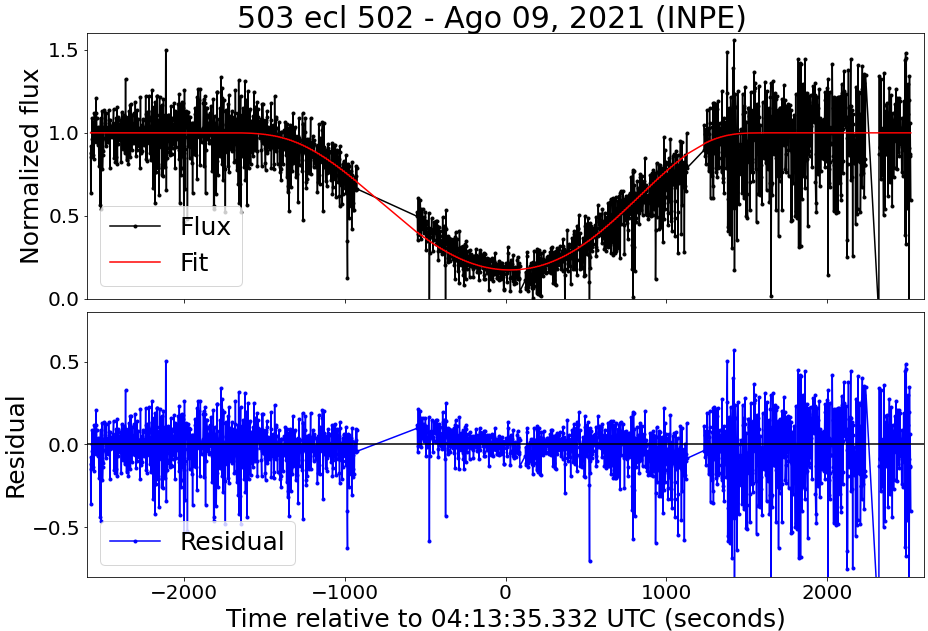}

    \end{minipage}
    \vfill
    \vspace{0.40cm}
    \centering
    \begin{minipage}{0.44\textwidth}
    \centering
    \includegraphics[width=1\textwidth]{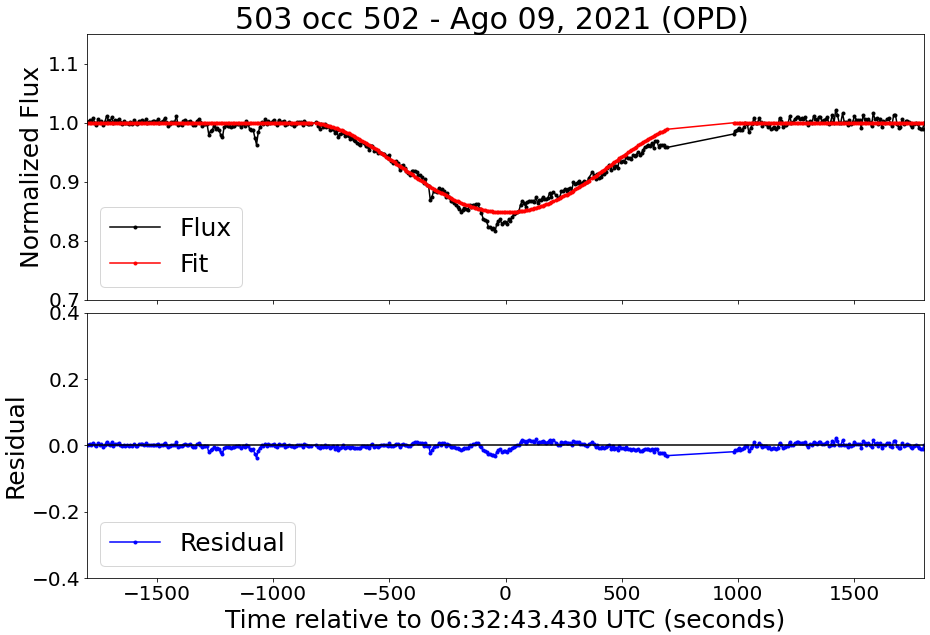}

    \end{minipage}
    \hspace{0.85cm}
    \centering
    \begin{minipage}{0.44\textwidth}
    \centering
    \includegraphics[width=1\textwidth]{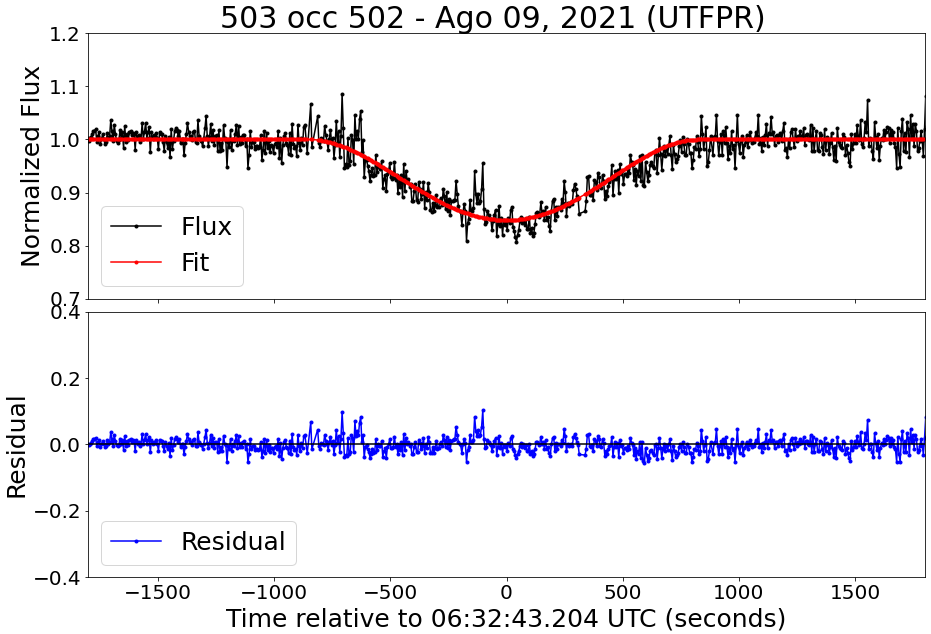}

    \end{minipage}
    \vspace{0.25cm}
    \caption{Fitting of the eight light curves of events between the Galilean satellites. The graph title indicates the satellites, event, date, and observer. The upper panel of each figure displays the fitting of the light curves, and the lower panel shows the residuals of the fitting.}
    \label{fig:04}
\end{figure*}



\subsection{Galilean satellite events}

Here we present the results of the events that involved only the Galilean satellites. The individual results for each parameter of interest can be seen in Table \ref{tab:04}. Table \ref{tab:05} shows the topocentric results for the positions $X$ and $Y$.

Table \ref{tab:04} presents, in the first three columns, the date, event type, and observer, as defined in Tables \ref{tab:01} and \ref{tab:02}. In the following columns, we show the results for the impact parameters ($S_0$), central instant ($t_0$), and relative velocity ($v_0$). Additionally, the deviations (O - C) between the observations and the JPL DE440 + JUP365 ephemerides are provided. The deviations are represented by $\Delta S_0$, $\Delta t_0$, and $\Delta v_0$. In the last two columns, we display the number ($N$) of used images and the normalized $\chi^2$. At the bottom of the table, the means and standard deviations (S. D.) are presented.

Table \ref{tab:05}, with the first three rows analogous to Table \ref{tab:04}, also presents the results in terms of topocentric distances $X$ and $Y$ between the pair of satellites involved in the event (occulting/eclipsing satellite minus occulted/eclipsed satellite), as well as the uncertainties ($\sigma X$, $\sigma Y$), the ephemeris offsets ($\Delta X$, $\Delta Y$) and the number of points used to fit the curves ($N$). The means and standard deviations for the parameters are provided at the bottom of the table, similarly to Table \ref{tab:04}.

Fig. \ref{fig:04} displays the plots with curve fitting for each event involving the Galilean satellites observed during the 2021 campaign. In the top panel of each plot, the red curve represents the model, and the black curve shows the normalized flux. In the lower panel, shown in blue, the resulting residual from the model's fit to the observations is presented.

The albedo ratio between Ganymede and Callisto could not be determined for the occultation that occurred on August 9th. This was because the satellites involved in the occultation were not sufficiently separated to measure the individual flux from both in the images obtained. Therefore, to avoid a significant loss of precision, the same albedo ratio values published by \cite{morgado2019} were used, where the albedo ratio between Ganymede and Callisto was 1.61.

According to Table \ref{tab:04}, the average errors achieved in our analyses were as follows: for the impact parameter, the average error was 4.7 $mas$; for the central instant, we obtained an average error of about 4.6 $mas$; and regarding the relative velocity, the average error was 0.2 $mas/s$. 

As for the ephemeris offsets, which represent the differences between the observed and predicted values, we obtained the following results: for the impact parameter, the mean offset was -5.6 $mas$ with a standard deviation of 5.1 $mas$; for the central instant, the mean offset was +0.8 $mas$ with a standard deviation of 2.21 $mas$; and for the relative velocity, the mean offset was +0.01 $mas/s$ with a standard deviation of 0.05 $mas/s$. All these values can be found in Table \ref{tab:04}. The accuracy of our analyses was 2.4 $mas$.

The error values are consistent with the dispersion of the ephemeris offsets, with no mean offsets greater than 1 sigma ($1\sigma$). Therefore, we can conclude that the JPL DE440 + JUP365 ephemerides are in good agreement with the observations from the 2021 campaign of mutual phenomena of the Galilean satellites, at the level of 5 $mas$.

\subsection{Eclipse of Thebe}

\begin{table}
    \centering
    \caption{Results of the eclipse of Thebe by Ganymede in 2021.}\label{tab:06}
    \begin{tabular}{lccc}
    \hline \hline
    
{Parameters} & {Ephemeris} & {Observed (error)} & {(O - C)} \\

  \hline 
  
$t_{0}$ (h:m:s)  & 04:05:18.271 & 04:05:19.491 (6.883s) & +11.64 $mas$ \\
$S_0$ ($mas$) & 816.50 & 820.67 (2.435) & +4.164 \\ 
$v_0$ ($mas/s$) & 8.044  & 9.547 (1.097) & +1.503 \\ 
$\sigma$ (O - C) & - & 0.279  & - \\ 
$N$ & - & 43.00 & - \\
$\chi^2$ & - & 0.731 & - \\ 

\hline 
\end{tabular}
\noindent \justifying{\textit{Note:} The time is given in UTC.}
\end{table}


\begin{table}
    \centering
    \caption{Results in $X$ e $Y$ for the eclipse of Thebe by Ganymede on July 18, 2021. \label{tab:07}}
    \begin{tabular}{ccccccc}
    \hline \hline
 {$X$} & {$Y$} & {$\sigma X$} & {$\sigma Y$} & {$\Delta X$} & {$\Delta Y$} & {$N$} \\

 {($mas$)} & {($mas$)} & {($mas$)} & {($mas$)} & {($mas$)} & {($mas$)} & {}
\\
\hline 

 +301,74 & -763,26 & 38,86 & 51,06 & -9,17 & +8,28 & 43 \\

\hline 

\end{tabular}

\end{table}

In the observations of the mutual phenomena in 2021, we were able to record, using the 1.6 m telescope at the Pico do Dias Observatory, a rare eclipse between the satellites Ganymede and the inner satellite Thebe, which occurred in the early morning of July 18, 2021. Performing astrometry of Thebe through mutual phenomena is a challenging task due to the proximity of the satellite to Jupiter (with a semi-major axis equivalent to 3.17 Jupiter radii), which often causes obstruction due to the scattered brightness of the planet.

To overcome the mentioned observational problems, the coronography technique is commonly applied. This technique has shown positive results for performing astrometry of the inner satellites, as it allows the separation of the satellite from Jupiter's scattered light \citep{christou2010, saquet2016, robert2017, morgado2019}.

However, we opted for a different approach and applied a technique of subtracting Jupiter's scattered light from the observational data by constructing a model of the planet's scattered flux (see Section \ref{sec:5}). This approach was innovative in the analysis of mutual phenomena, and the results were very promising, making it feasible and opening new avenues for the analysis of mutual phenomena involving Jupiter's inner satellites.

Note that the first observation of a mutual phenomenon involving the inner satellite Thebe was reported by \cite{saquet2016}, who published the results of an eclipse of Thebe by Callisto observed during the 2014/2015 campaign. Thanks to our observational efforts and analysis approach, we were able to provide the second observation of a mutual phenomenon involving Thebe.

The photometry of this dataset was also performed using the PRAIA package \citep{assafin2023}. However, for this event, during the reduction process, we manually defined the photometric aperture to achieve the best signal-to-noise ratio. It is worth noting that for these data, we do not have a photometric calibrator available (see Table \ref{tab:03}), and therefore, systematic variations in the light curve of this event are related to changes in the night's weather conditions.

The simulations and light curve fitting of the eclipse of Thebe were conducted following the approach described in Section \ref{sec:5}. However, due to Thebe's irregular shape (triaxial diameter of 116 × 98 × 84 km), in our simulations, we considered the satellite as a sphere with a radius of 49.3 km, with an uncertainty of 4 km \citep{thomas1998}. This choice does not affect the accuracy of the measurements, as the relative velocity of the event involving Thebe is about 9.5 $mas/s$ ($\sim$ 28 km/s), and with a temporal resolution of 5 seconds in our measurements, coupled with the resolution of our observations ($\sim$ 140 km), the real shape of the satellite becomes indistinguishable from the approximate shape.

The results obtained from the observation of the eclipse of Thebe are compiled in Table \ref{tab:06}. The parameters listed in the first column are: the central instant $t_0$; the impact parameter $S_0$; the relative velocity $v_0$; the standard deviation $\sigma$ (O - C); the number $N$ of data points used in the fitting; and the value of the minimum chi-squared per degree of freedom found with the curve fitting. In the second column, we provide the values predicted by the JPL DE440 + JUP365 ephemerides; in the third column, we display the results obtained from the curve fitting; and finally, in the fourth column, we present the offsets between the observations and the ephemerides (O - C). As before, the JPL DE440 + JUP365 ephemerides were used in this analysis.

\begin{figure}
    \centering
	\includegraphics[width=\columnwidth]{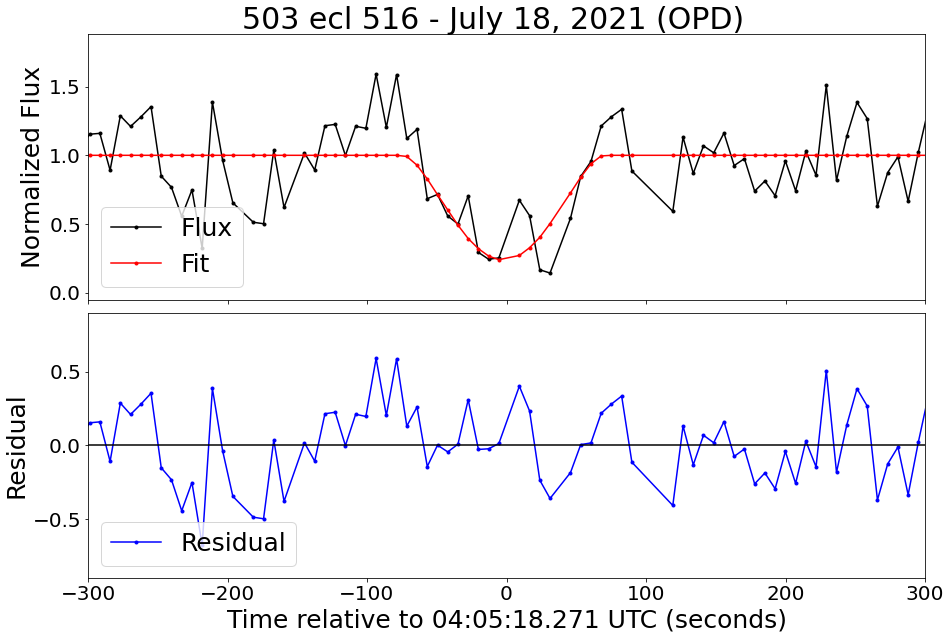}
    \caption{ Light curve fit of an eclipse between Ganymede (503) and Thebe (514) observed on 1.6 m telescope on Pico dos Dias observatory (OPD) in July 18, 2021.}
    \label{fig:05}
\end{figure}

Fig. \ref{fig:05} displays the light curve fitting of Thebe. Similar to plots of Fig. \ref{fig:04}, the upper panel shows the black curve representing the observation, and the red curve represents the fitted model. In the lower panel, the blue curve shows the residual of the fit.

The uncertainty achieved for the impact parameter, as indicated in Table \ref{tab:06}, reached 2.4 $mas$. For the relative velocity, the error was 1.1 $mas/s$, and for the central instant, it was 6.7 seconds $mas$ between the observed and predicted values. The offsets were +4.2 $mas$ for the impact parameter, 1.5 $mas/s$ for the relative velocity, and 11.6 $mas$ for the central instant.

Table \ref{tab:07} displays the results concerning the right ascension ($\alpha$) and declination ($\delta$) directions for the positions ($X$, $Y$). The information contained in Table \ref{tab:07} is analogous to that of Table \ref{tab:05}. It is noteworthy that the errors $\sigma X$ and $\sigma Y$ in the topocentric positions were, respectively, 39 $mas$ ($\sim$ 117 km) and 51 $mas$ ($\sim$ 153 km).

\section{Conclusion}\label{sec:9}

In this study, we present the results of 9 light curves obtained from the observation of 7 different mutual events of Jupiter in 2021. The observed events involved the Galilean satellites, and one particular event included the inner satellite Thebe. The observations were conducted at the Pico dos Dias Observatory and by collaborators in the south and southeast regions of Brazil. Telescopes with apertures ranging from 0.20m to 1.60m were used for the observations.

The analysis method we employed involves applying simulation routines and curve fitting techniques described in Sections \ref{sec:5} and \ref{sec:6}, as published in \cite{morgado2019}. During our fitting process, we encountered average offset values that are below 1 sigma ($1\sigma$). The minimum average precision obtained for events involving only the Galilean satellites was 2.8 $mas$ ($\sim$ 8.4 km), as shown in Table \ref{tab:05}. For the Thebe eclipse, we achieved a minimum precision of 38.8 $mas$ ($\sim$ 116.7 km). Based on these results, it can be concluded that our modeling aligns well with the observational data, demonstrating a strong agreement between the model predictions and actual observations.

The results derived from our analyses, combined with the extensive set of observations of this system, are significant for quantifying variations in the orbits and velocities of the satellites. These measurements can be utilized to enhance the uncertainties associated with the orbits of the Galilean satellites and Jupiter's inner satellite, Thebe.

More precise ephemerides of Jupiter's satellite orbits, especially the Galilean satellites, are crucial for the planning of space missions aimed at this system, as well as the optimization of ongoing missions. In this context, the Europa Clipper mission\footnote{Website: \href{https://europa.nasa.gov/}{https://europa.nasa.gov/}}, scheduled to launch in 2024, and the JUpiter ICy moons Explorer (JUICE) probe\footnote{Website: \href{https://www.esa.int/Science_Exploration/Space_Science/Juice}{https://www.esa.int/Science\_Exploration/Space\_Science/Juice}}, launched in April 2023 with an expected arrival at the Jovian system in 2031, stand out. Both probes will derive precise space astrometry with flies by all Galilean moons, except Io \citep{fayolle2023}. This further enhances the importance of our study, as 3 out of the 5 independent events here analysed involve Io.

Furthermore, periodic measurements of the orbits of the Galilean satellites enable the study of low-intensity effects that are less explored, such as resonant effects and tidal dissipation \citep{lainey2009}.

Through the observation and data processing techniques mentioned in Sections \ref{sec:02} and \ref{sec:03}, we were able to observe the second mutual phenomenon involving the inner satellite Thebe. This was first reported by \cite{saquet2016} based on observations from the 2014-2015 campaign. This approach is particularly intriguing as it paves the way for future studies of mutual phenomena involving Jupiter's inner satellites or even the reevaluation of observations from previous campaigns.

In the context of this work, we encourage the continuous observation of mutual phenomena occurring among the moon of giant planets. The next mutual phenomena of Jupiter is expected to take place between 2026 and 2027, and an observational campaign for these events will be organized for that occasion. Additionally, it's worth emphasizing the significance of observing the 2024-2026 campaign for the mutual phenomena of Saturn. After the conclusion of the Cassini mission, the Saturnian system requires new astrometric measurements to maintain the precision of its satellite ephemerides. Predictions for the upcoming mutual phenomena of Jupiter and Saturn can be consulted on the IMCCE website\footnote{Website: \href{http://nsdb.imcce.fr/multisat/nssephme.htm}{http://nsdb.imcce.fr/multisat/nssephme.htm}} \citep{arlot2019}.

\section*{Acknowledgements}

We express our sincere gratitude to the group of observers who collaborated to make this work possible. We also extend our thanks to the team at the Pico dos Dias Observatory for their support and technical assistance during the observations, which took place amidst the crisis triggered by the COVID-19 pandemic. M.A. thanks CNPq grants 427700/2018-3, 310683/2017-3 and 473002/2013-2. B.E.M. thanks CNPq grant 150612/2020-6. F.B.R. acknowledges CNPq grant number 314772/2020-0. R.V.M thanks grant CNPq 307368/2021-1. I.J.L acknowledges São Paulo Research Foundation (FAPESP) for financial support under grant 2015/24383-7 and  2013/26258-4 This work was based on observations primarily conducted at the Pico dos Dias Observatory, managed by the Laboratório Nacional de Astrofísica (LNA), Itajubá-MG, Brasil. We also received observations from collaborators in Brazil's southern and southeastern regions. This study was funded by the Coordenação de Aperfeiçoamento de Pessoal de Nível Superior - Brasil (CAPES), Finance Code 001.

\section*{Data Availability}

The data that support the results and plots in this paper and other findings of this study are available at the NSDB database \href{http://nsdb.imcce.fr/obsphe/obsphe-en/}{http://nsdb.imcce.fr/obsphe/obsphe-en/}, or directly available from the corresponding author upon reasonable request.



\bibliographystyle{mnras}
\bibliography{bibliography}



\appendix




\bsp	
\label{lastpage}
\end{document}